\documentclass[preprint,showpacs,preprintnumbers,amsmath,amssymb]{revtex4}
\usepackage{graphicx}
\usepackage{epsfig}
\usepackage{bm}
\usepackage{amsfonts}
\usepackage{subfigure}
\usepackage{multirow}
\usepackage{float}
\usepackage{color}
\usepackage{ulem}
\usepackage{todonotes}
\usepackage{makecell}
\usepackage{enumitem}
\usepackage{amsmath}
\usepackage{hyperref}
\usepackage{amssymb}
\usepackage{caption}
\usepackage{verbatim}
\graphicspath{{./figs/}}
\hypersetup{colorlinks=True,citecolor=blue}
\def\thickhline{\noalign{\hrule height1.9pt}}
\usepackage[font=small]{caption}   
\begin{document}

\title{Revisited apparent horizon entropy and GSL in modified gravity}
\author{ Soma Heydari\footnote{s.heydari@uok.ac.ir}, Parastoo Askari\footnote{p.askari.1167@gmail.com}, and Kayoomars Karami\footnote{kkarami@uok.ac.ir}}
\address{\small{Department of Physics, University of Kurdistan, Pasdaran Street, P.O. Box 66177-15175, Sanandaj, Iran}}
\date{\today}
 \begin{abstract}
This work presents a universal and revisited formalism for the entropy of the apparent horizon in modified gravity to investigate the validity of the Generalized Second Law (GSL) of thermodynamics. This revisited horizon entropy is constructed directly from the modified Friedmann equations in a Friedmann-Robertson-Walker (FRW) universe.
The resulting entropy relation contains, beside the standard Bekenstein-Hawking term, an additional integral contribution that encodes the effective energy density and pressure generated by deviations from general relativity.  Using this universal entropy formula, a compact expression for the GSL is derived. This formalism is then applied to some viable $f(T)$ and $f(R)$ gravity models, in order to re-evaluate the validity of the GSL as a function of redshift. The analysis demonstrates that including the integral term in the revisited entropy can relatively improve the late-time validity of the GSL for some of these models while living others unchanged, thereby reinforcing the profound connection between thermodynamics and gravity.
\end{abstract}
\maketitle

\section{Introduction}

The evidence for the current accelerated expansion phase of the universe has been provided by joint observational data from Type Ia supernovae (SNIa) \cite{Riess}, the Cosmic Microwave Background (CMB) \cite{Spergel,Komatsu:2011}, and Large-Scale Structure (LSS) \cite{Eisenstein:2005}. Expounding this accelerated expansion phase quickly became the most significant challenges in modern cosmology, which motivates two main theoretical approaches. In the first one, a mysterious component with negative pressure, namely Dark Energy (DE) is introduced within the framework of Einstein General Relativity (GR) \cite{Padmanabhan,Copeland}. In the second one, modifications to GR are considered, which give rise to theories of modified gravity \cite{Sotiriou:2010,Nojiri:2011,De Felice:2010,Nojiri:5}.

Among the most studied modified gravity theories are $f(R)$ gravity, which generalizes the Einstein-Hilbert action to an arbitrary function of the Ricci scalar $R$, and $f(T)$ gravity, built upon a teleparallel formulation of gravity where the gravitational interaction is described by torsion instead of curvature \cite{bengochea,linder}. These theories offer a gravitational alternative to DE, with the aim of unifying the early-time inflationary epoch with the late-time cosmic acceleration under a single theoretical construct.

Setting up a deep connection between gravity and thermodynamics has been a favorable area of progress in theoretical cosmology \cite{Sebastiani:2023,Sheykhi:2025,Luciano:2025,Nojiri:2022,Nojiri:1,Nojiri:2,Nojiri:3,Nojiri:4,Nojiri:6}. This originated from the discovery of black hole thermodynamics \cite{Bardeen:1973,Bekenstein} and was later extended to cosmological horizons. Fundamental studies like \cite{Jacobson:1995} have established the increase of black hole entropy in higher-curvature gravity and extended the second law of black hole thermodynamics to generalized gravitational theories. It has been shown that the Friedmann equations, which govern the dynamics of the universe, can be derived from the first law of thermodynamics, the Clausius relation. To do so, the Clausius relation must be applied to the apparent horizon entropy \cite{Cai:2005}. This thermodynamic interpretation of gravity has been successfully expanded to various  modified gravity models \cite{Akbar:2007,Sheykhi:2009,Karami:2011,Karami:2012,Karami:2013,
Karami-epl:2012,Mitra:2014,Asadzadeh:2016,
Beigmohammadi:2020,Panda:2024}.

A critical thermodynamic test for any physical system is the validity of the Generalized Second Law (GSL) of thermodynamics. This law posits that the total entropy of the universe, consisting of the entropy of matter and fields enclosed by a horizon and the entropy of the horizon itself, must not decrease over time. The GSL has been extensively explored in various cosmological scenarios \cite{Izquierdo:2006,Zhou:2007,Kirklin:2025,Dhivakar:2024}. The conventional thermodynamic approach in modified gravity presumes a direct modification of the Bekenstein-Hawking entropy, such as $S_{\rm A}=A f_{\rm T}/4{G}$ in $f(T)$ gravity \cite{Miao:2011} and $S_{\rm A}=A f_{\rm R}/4{G}$ in $f(R)$ gravity \cite{Wald}. However, such definitions can yield inconsistencies or require specific limiting conditions to hold.

In order to address this issue, in this paper a universal and revisited formalism is derived for the apparent horizon entropy, applicable to a broad class of modified gravity theories. In this formulation, a general expression for the apparent horizon entropy is derived from the modified Friedmann equations of modified gravity. The revisited relation for apparent horizon entropy incorporates contributions from the standard Bekenstein–Hawking term plus an integral term that depends on the effective energy density and pressure resulting from the modifications to gravity. We demonstrate that this revisited entropy ensures the equivalence between the Clausius relation and the gravitational field equations. On this basis, a general form of the GSL is obtained for modified gravity, expressed as a simple condition on the combined evolution of horizon and matter entropies. Subsequently, this formalism is applied  to re-examine the validity of the GSL for some viable $f(T)$ and $f(R)$ gravity models. For each case, the horizon entropy, matter entropy, and total entropy are computed as functions of redshift, and the impact of the integral correction term is quantified by comparing the revisited entropy prescription with the standard one that omits this term.

The body of this paper is organized as follows: In Section \ref{sec:base}, the universal relation for horizon entropy in modified gravity is introduced. Section \ref{sec:GSL} is devoted to obtaining a universal relation for GSL in modified gravity. In Sections \ref{sec:ft} and \ref{sec:fr}, we apply our revisited formalism to re-examine the GSL in $f(T)$ and $f(R)$ gravity, respectively. Finally, we summarise the results of the paper in Section \ref{sec:con}.
\section{Apparent horizon entropy in modified gravity}{\label{sec:base}}

In this section,  we introduce a universal equivalency between thermodynamics and gravity through a key relation for the horizon entropy in modified gravity. Considering a non-flat FRW universe,
the reformed Friedmann equations in modified gravity models are given as follows
\begin{eqnarray}
  H^2+\frac{K}{a^2} &=& \frac{8\pi G}{3}\rho_{\rm t},\label{f1}\\
  \dot{H}-\frac{K}{a^2}&=& -4\pi G(\rho_{\rm t}+p_{\rm t}),\label{f2}
\end{eqnarray}
where the curvature constant $K$ can be $0$, $1$, or $-1$, corresponding to a flat, closed, or open universe, respectively. Also $H\equiv\dot{a}/a$ is the Hubble parameter, where $a$ is the scale factor, and the dot denotes differentiation with respect to cosmic time $t$. In addition, $\rho_{\rm t}=\rho_{\rm m}+\rho_{\rm e}$ is the total energy density and $p_{\rm t}=p_{\rm m}+p_{\rm e}$ is the total pressure of the fluid inside the horizon. Here, $\rho_{\rm m}$ and $p_{\rm m}$
indicate the energy density and pressure of matter, respectively. Also $\rho_{\rm e}$ and $p_{\rm e}$ correspond to the effective energy density and pressure arising from the modification to gravity. Moreover, the energy conservation laws are read by
\begin{equation}\label{clm}
\dot{\rho}_{\rm m}+3H(\rho_{\rm m}+p_{\rm m})=0,
\end{equation}
\begin{equation}
\dot{\rho}_{\rm e}+3H(\rho_{\rm e}+p_{\rm e})=0.\label{cle}
\end{equation}
Notably, for $\rho_{\rm e}=0=p_{\rm e}$, all the above equations return to their Einstein gravity counterparts.

Considering a spatially non-flat FRW metric, the apparent horizon takes the following form
\begin{equation}
\tilde{r}_{\rm A}=\left(H^2+\frac{K}{a^2}\right)^{-1/2}.\label{rA}
\end{equation}
In this case, the Hawking temperature on the apparent horizon is given by \cite{Cai:2005}
\begin{equation}
T_{\rm A}=\frac{1}{2\pi \tilde{r}_{\rm
A}}\left(1-\frac{\dot{\tilde{r}}_{\rm A}}{2H\tilde{r}_{\rm A}}
\right).\label{TA1}
\end{equation}
In order to demonstrate the equivalence between thermodynamics and modified gravity, we can derive a universal equation for the horizon entropy. This approach combines the Clausius relation with second modified Friedmann equation, Eq. (\ref{f2}). The Clausius relation is given by
\begin{equation}\label{Clausius}
\delta Q=-{\rm d}E=T_{\rm A}{\rm d}S_{\rm A},
\end{equation}
where ${\rm d}E$ is the energy traversing the apparent horizon during an infinitesimal time interval ${\rm d}t$.
Crucially, the apparent horizon radius is momentarily assumed to be static within this interval, i.e., $\dot{\tilde{r}}_{\rm A}=0$ \cite{Cai:2009}. This assumption is used only at that specific instant and simplifies the Hawking temperature Eq. (\ref{TA1}) to
\begin{equation}\label{TA2}
  T_{\rm A}\Big|_{\dot{\tilde{r}}_{\rm A}=0}=\frac{1}{2\pi \tilde{r}_{\rm A}}.
\end{equation}
The Clausius relation (\ref{Clausius}), yields
\begin{equation}\label{clu}
{\rm d}S_{\rm A}=-\frac{{\rm d}E}{T_{\rm A}}=-2 \pi \tilde{r}_{\rm A}{\rm d}E.
\end{equation}
Using $E=\rho_{\rm m} V$ where ${\rm V}=\frac{4}{3}\pi \tilde{r}_{\rm A}^3$ is the volume enclosed by the apparent horizon, and the continuity Eq. (\ref{clm}), one can find
\begin{equation}\label{dE}
  -{\rm d}E=4\pi\tilde{r}^3_{\rm A}H(\rho_{\rm m}+p_{\rm m}){\rm d}t.
\end{equation}
From the second Friedmann Eq. (\ref{f2}), we have
\begin{equation}\label{rom+pm}
  \rho_{\rm m}+p_{\rm m}=\frac{-1}{4\pi G}\left(\dot{H}-\frac{K}{a^2}\right)-(\rho_{\rm e}+p_{\rm e}).
\end{equation}
Substituting this result into the energy flow expression (\ref{dE}) gives
\begin{equation}\label{de}
  -{\rm d}E=-\frac{H\tilde{r}^3_{\rm A}}{G}\left(\dot{H}-\frac{K}{a^2}\right)-4\pi H\tilde{r}^3_{\rm A} (\rho_{\rm e}+p_{\rm e}).
\end{equation}
Substituting Eq. (\ref{de}) into (\ref{clu}) and then integrating yields the horizon entropy
\begin{equation}\label{clu2}
 S_{\rm A}=\int-\frac{2\pi H\tilde{r}^4_{\rm A}}{G}\left(\dot{H}-\frac{K}{a^2}\right){\rm d}t-8\pi^2\int H\tilde{r}^4_{\rm A} (\rho_{\rm e}+p_{\rm e}){\rm d}t.
\end{equation}
The first integral can be shown to be the Bekenstein-Hawking entropy, $A/(4G)$, where $A=4\pi \tilde{r}_{\rm A}^2$ denotes the area of the apparent horizon. To show this, we use the time derivative of the apparent horizon radius from Eq. (\ref{rA})
\begin{equation}\label{rdot}
  \dot{\tilde{r}}_{\rm A}=-H \tilde{r}^3_{\rm A}\left(\dot{H}-\frac{K}{a^2}\right)=4\pi G H\tilde{r}_{\rm A}^3(\rho_{\rm t}+p_{\rm t}),
\end{equation}
where the second equality follows from the second Friedmann Eq. (\ref{f2}). Using the first equality of Eq. (\ref{rdot}), the first integral term in Eq. (\ref{clu2}) indeed yields the Bekenstein-Hawking entropy term. Ergo, for a spatially non-flat FRW universe filled with the matter and bounded by the dynamical apparent horizon, the revisited horizon entropy is given by the universal expression
\begin{equation}
S_{\rm A}=\frac{A}{4G}-8\pi^2\int H\tilde{r}_{\rm A}^4(\rho_{\rm e}+p_{\rm e}){\rm d}t.\label{sa}
\end{equation}
For any specific model of modified gravity, the horizon entropy $S_{\rm A}$ can be computed from Eq. (\ref{sa}) once the effective energy density $\rho_{\rm e}$ and pressure  $p_{\rm e}$ are known.
Note that in the case where $\rho_{\rm e}=0=p_{\rm e}$, the second term in Eq. (\ref{sa}) vanishes, and the Bekenstein-Hawking entropy $S_{\rm A}={\rm A}/(4G)$ for Einstein gravity is recovered. The second term, therefore, represents the entropy correction arising from the modifications to general relativity.

In modified gravity, the thermodynamic interpretation can also be formulated in a non‑equilibrium form, where an entropy production term $d_{i}S$ appears to involve bulk viscosity entropy production e.g., $dS=\frac{\delta Q}{T}+d_{i}S$  \cite{Eling:2006}. In the present work we follow an alternative but common strategy. We keep the equilibrium Clausius relation Eq. (\ref{Clausius}) and instead revisit the horizon entropy functional so that it is constructed directly from the modified Friedmann equations \cite{Cai:2005,Akbar:2007}. In this sense, the integral correction term in Eq. (\ref{sa}) plays the role of encoding the deviation from GR and restores consistency between the thermodynamic relation and the gravitational dynamics within our FRW setup.

\section{Generalized Second Law of Thermodynamics in Modified Gravity }{\label{sec:GSL}}

In this section, a general, revisited form of the GSL of thermodynamics is investigated in the context of modified gravity. The GSL states that the total entropy, the sum of the horizon entropy and the entropy of matter within the horizon, must never decrease with time \cite{Cai:2005}.

First, we calculate the rate of change of the horizon entropy, $\dot{S}_{\rm A}$. To do this, using Eqs. (\ref{rA})-(\ref{TA1}) and the time derivative of Eq. (\ref{sa}), we find the term $T_{\rm A}\dot{S}_{\rm A}$ for any given modified gravity model as follows
\begin{equation}
T_{\rm A}\dot{S}_{\rm A}=\left(1-\frac{\dot{\tilde{r}}_{\rm A}}{2H\tilde{r}_{\rm A} }
\right)\left[\frac{\dot{\tilde{r}}_{\rm A}}{G}-4\pi H\tilde{r}_{\rm A}^3(\rho_{\rm e}+p_{\rm e})\right].\label{tsa}
\end{equation}
Substituting  $\dot{\tilde{r}}_{\rm A}$  from the second equality of Eq. (\ref{rdot}) into (\ref{tsa}) yields
\begin{equation}
T_{\rm A}\dot{S}_{\rm A}=4\pi H\tilde{r}_{\rm A}^3(\rho_{\rm m}+p_{\rm m})\big[1-2\pi G\tilde{r}_{\rm A}^2(\rho_{\rm t}+p_{\rm t})\big].\label{tsa2}
\end{equation}
Next, we calculate the contribution from the matter entropy using the Gibbs equation \cite{Izquierdo:2006}
\begin{equation}
T_{\rm A}{\rm d}S_{\rm m}={\rm d}E+p_{\rm m}{\rm d}{\rm V}.\label{Tdsm1}
\end{equation}
Taking time derivative of Eq. (\ref{Tdsm1}), and using $E=\rho_{\rm m} V$ as well as the continuity Eq. (\ref{clm}), one can find the entropy contribution from the matter within the apparent horizon as follows
\begin{equation}
T_{\rm A}\dot{S}_{\rm m}=4\pi\tilde{r}_{\rm A}^2(\rho_{\rm m}+p_{\rm m})(\dot{\tilde{r}}_{\rm A}-H\tilde{r}_{\rm A}).\label{Tdsm2}
\end{equation}
Substituting  $\dot{\tilde{r}}_{\rm A}$ from the second equality of Eq. (\ref{rdot}) into (\ref{Tdsm2}) gives
\begin{equation}
T_{\rm A}\dot{S}_{\rm m}=4\pi\tilde{r}_{\rm A}^2(\rho_{\rm m}+p_{\rm m})\Big[4\pi H\tilde{r}^3_{\rm A}G(\rho_{\rm t}+p_{\rm t})-H\tilde{r}_{\rm A}\Big].\label{Tdsmf}
\end{equation}
To establish the revisited GSL of thermodynamics, we add two contributions from Eqs. (\ref{tsa2}) and (\ref{Tdsmf}) as follows
\begin{equation}
T_{\rm A}\dot{S}_{\rm tot}=8\pi^2GH\tilde{r}^5_{\rm A}(\rho_{\rm m}+p_{\rm m})(\rho_{\rm t}+p_{\rm t})=\frac{1}{2}A(\rho_{\rm m}+p_{\rm m}) \dot{\tilde{r}}_{\rm A},\label{gsl}
\end{equation}
where $\dot{S}_{\rm tot}=\dot{S}_{\rm m}+\dot{S}_{\rm A}$ and the second equality has been obtained from Eq. (\ref{rdot}). Note that in the case of GR, we have $\rho_{\rm e}=0=p_{\rm e}$ and $\rho_{\rm t}+p_{\rm t}=\rho_{\rm m}+p_{\rm m}$ and consequently Eq. (\ref{gsl}) reads
\begin{equation}
T_{\rm A}\dot{S}_{\rm tot}=8\pi^2GH\tilde{r}^5_{\rm A}(\rho_{\rm m}+p_{\rm m})^2\geq 0,
\end{equation}
which shows that the GSL is always satisfied in GR. Equation (\ref{gsl}) allows us to analyze the  GSL of thermodynamics for any given model of modified gravity in which the condition  $T_{\rm A}\dot{S}_{\rm tot}\geq 0$ must hold. In the next section, we will examine this revisited GSL of thermodynamics for some viable $f(T)$ and $f(R)$ gravity models.

It is worth noting that general proofs of the GSL, such as those discussed in \cite{Wall:2012,Sarkar:2013}, often rely on the assumption that matter fields satisfy specific energy conditions, most notably the Null Energy Condition (NEC), i.e., $(\rho_{\rm e}+p_{\rm e})\ge0$. In the context of modified gravity,  the field equations are often expressed in an Einstein-like form by introducing effective energy density $\rho_{\rm e}$ and pressure $p_{\rm e}$ for the geometric sector. These effective contributions need not satisfy the standard energy conditions. For the effective Strong Energy Condition (SEC) to be satisfied, it requires both $(\rho_{\rm e}+p_{\rm e})\ge0$ and $(\rho_{\rm e}+3p_{\rm e})\ge0$. Indeed, the violation of the SEC, characterized by $(\rho_{\rm e}+3p_{\rm e})\le0$, is a general requirement for any theory describing the late-time accelerated expansion of the universe. Consequently, the general expression in Eq. (\ref{gsl}) should be regarded not as a fundamental proof of the GSL for arbitrary states, but as a cosmological consistency test. The thermodynamic viability of a specific gravity model must therefore be evaluated by investigating the evolution of the total entropy rate alongside the status of the effective energy conditions, assuming the physical matter sectors
$\rho_{\rm m}$ and $p_{\rm m}$, obey the standard conservation law (\ref{clm}).
\section{$f(T)$ gravity}{\label{sec:ft}}
Here, we obtain the apparent horizon entropy and examine the GSL in $f(T)$ gravity. We begin by reviewing the basic formulation of the $f(T)$ modified
teleparallel gravity with the following action  \cite{Ferraro:2007,Ferraro:2008}
\begin{equation}
{\cal S}=\frac{1}{16\pi G}\int {\rm d}^4
x~e~\Big[f(T)+{\cal L}_{\rm m}\Big],\label{action}
\end{equation}
where $e\equiv{\rm det}(e^i_{\mu})$ with $e^i_{\mu}$ being the
vierbein field that serves as the dynamical object in teleparallel gravity. Also $T$ is the torsion scalar, and ${\cal L}_{\rm m}$ is the matter Lagrangian density.

For simplicity, we consider a spatially flat FRW universe ($K=0$) which is compatible with current observations. In this case, the dynamical apparent horizon (\ref{rA}) is same as the Hubble
horizon, i.e. $\tilde{r}_{\rm A}=H^{-1}$. Then, within the framework of $f(T)$ gravity, the effective energy density $\rho_{\rm e}$ and pressure $p_{\rm e}$ arising from torsion contributions to the reformed Friedmann equations (\ref{f1}) and (\ref{f2}) are given by \cite{Wu:2010,Wu:2011,Wei:2011,Karami:2013b}
\begin{equation}
\rho_{\rm e}=\frac{1}{16\pi G}\Big(2Tf_{\rm T}-f-T\Big),\label{roeT}
\end{equation}
\begin{equation}
p_{\rm e}=-\frac{1}{16\pi
G}\Big[-8\dot{H}Tf_{\rm TT}+\Big(2T-4\dot{H}\Big)f_{\rm T}-f+4\dot{H}-T\Big],\label{peT}
\end{equation}
\begin{equation}
T=-6H^2.\label{T}
\end{equation}
Here, $f_{\rm T}$ and $f_{\rm TT}$  denote the first and second derivatives of $f(T)$ with respect to the torsion scalar $T$, respectively. From Eqs. (\ref{roeT}) and (\ref{peT}), it is clear that for $f(T)=T$, we have
$\rho_{\rm e}=0=p_{\rm e}$. In this limit, the Friedmann Eqs. (\ref{f1}) and (\ref{f2})  reduce to their standard forms in GR.
Substituting Eqs. (\ref{roeT}) and (\ref{peT}) into the reformed Friedmann equations (\ref{f1}) and (\ref{f2}) for pressureless matter $(p_{\rm m}=0)$ gives
\begin{equation}
\dot{H}=-\frac{4\pi G\rho_{\rm m}}{f_{\rm T}+2Tf_{\rm TT}}.\label{Hdot}
\end{equation}
Furthermore, substituting Eqs. (\ref{roeT}) and (\ref{T}) into Eq. (\ref{f1}) allows us to obtain the matter energy density, $\rho_{\rm m}$
\begin{equation}
\rho_{\rm m}=\frac{1}{16\pi G}\big(f-2Tf_{\rm T}\big). \label{romT}
\end{equation}
It is important to note that $\rho_{\rm m}$ here represents the physical matter fluid, expressed via the first Friedmann equation in terms of $T=-6H^2$. Although written in terms of geometric quantities compatible with the given $f(T)$ background, it satisfies the conservation law Eq. (\ref{clm}) for pressureless matter $(p_{\rm m}=0)$ through
\begin{equation}\label{rhoma}
  \rho_{\rm m}=\rho_{\rm m_0}a^{-3}.
\end{equation}
 Using Eqs. (\ref{Hdot}) and (\ref{romT}) one can obtain
\begin{equation}
\dot{\tilde{r}}_{\rm
A}=-\dot{H}/H^2=-\frac{3}{2T}\left(\frac{f-2Tf_{\rm T}}{f_{\rm T}+2Tf_{\rm TT}}\right),\label{rdotA}
\end{equation}
\begin{equation}
\dot{T}=-12H\dot{H}=3H\left(\frac{f-2Tf_{\rm T}}{f_{\rm T}+2Tf_{\rm TT}}\right).\label{Tdot}
\end{equation}

\subsection{Apparent horizon entropy in $f(T)$ gravity}
In order to find the apparent horizon entropy in $f(T)$ gravity, we first calculate the term $\rho_{\rm e}+p_{\rm e}$ by adding Eqs. (\ref{roeT}) and (\ref{peT}) as
\begin{equation}\label{rot+pt}
  \rho_{\rm e}+p_{\rm e}=\frac{\dot{H}}{4\pi G}\big(2Tf_{\rm TT}+f_{\rm T}-1\big).
\end{equation}
Substituting this expression into Eq. (\ref{sa}) gives
\begin{equation}
S_{\rm A}=\frac{A}{4G}+\frac{1}{4G}\int\dot{A}\big(2Tf_{\rm TT}+f_{\rm T}-1\big){\rm d}{t},\label{sat1-1}
\end{equation}
where we have used $\dot{H}=-H^2\dot{\tilde{r}}_{\rm A}$ and $\dot{A}=8\pi\tilde{r}_{\rm A}\dot{\tilde{r}}_{\rm A}$.

Next, we apply integration by parts to the integral in Eq. (\ref{sat1-1}). Letting $\int v{\rm d}u=uv-\int u{\rm d}v$ for ${\rm d}u\equiv\dot{A}{\rm d}t$ and $v\equiv\left(2Tf_{\rm TT}+f_{\rm T}-1\right)$, we obtain
\begin{equation}
S_{\rm A}=\frac{A}{4G}+\frac{A}{4G}\big(2Tf_{\rm TT}+f_{\rm T}-1\big)-\frac{1}{4G}\int\big(\dot{f}_{\rm T}+ 2\dot{T}f_{\rm TT}+2T\dot{f}_{\rm TT}\big)A{\rm d}t.\label{sat1-2}
\end{equation}
Using $\dot{f}_{\rm T}=\dot{T}f_{\rm TT}$, Eq. (\ref{sat1-2}) simplifies to
\begin{equation}
S_{\rm A}=\frac{Af_{\rm T}}{4G}+\frac{ATf_{\rm TT}}{2G}-\frac{3}{4G}\int A\dot{T}f_{\rm TT}{\rm d}t-\frac{1}{2G}\int AT{\rm d}f_{\rm TT}.\label{sat1-3}
\end{equation}
Since $AT=4\pi \tilde{r}_{\rm A}^2 (-6H^2)=-24 \pi$, hence the second and last terms of above equation will canceled each other and subsequently Eq. (\ref{sat1-3}) reads
 \begin{equation}
S_{\rm A}=\frac{Af_{\rm T}}{4G}+\alpha\int f_{\rm TT}~{\rm d}\ln T,\label{sat2}
\end{equation}
where $\alpha\equiv18\pi/G$.
Note that for $f(T)=T$, Eq. (\ref{sat2}) reduces to the well-known Bekenstein-Hawking entropy-area relation $S_{\rm A}=A/(4G)$, as expected in GR. Notably, the apparent horizon entropy in $f(T)$ gravity is often given in the simplified form $S_{\rm A}=A f_{\rm T}/(4G)$, which assumes $f_{\rm TT}\ll 1$  (or $\alpha=0$) \cite{Karami:2012}. In this work, we use the complete form of Eq. (\ref{sat2}), including the integral term, to compute the apparent horizon entropy for some viable $f(T)$ gravity models.

\subsection{GSL in $f(T)$ gravity}
We now analyze the GSL Using the modified horizon entropy from Eq. (\ref{sat2}). Taking the time derivative of Eq. (\ref{sat2}) and using the expression for $T_{\rm A}$ from Eq. (\ref{TA1}), we find
\begin{equation}
T_{\rm A}\dot{S}_{\rm A}=\frac{1}{2\tilde{r}_{\rm
A}G}\left(1-\frac{\dot{\tilde{r}}_{\rm A}}{2H\tilde{r}_{\rm A}}
\right)\Big[2\dot{\tilde{r}}_{\rm A}\tilde{r}_{\rm A}f_{\rm T}+\tilde{r}_{\rm A}^2\dot{T}f_{\rm TT}+\alpha(G/\pi)\dot{T}f_{\rm TT}T^{-1}\Big].
\label{saft-1}
\end{equation}
Substituting the expressions  $\dot{\tilde{r}}_{\rm A}$ from Eq. (\ref{rdotA}) and $\dot{T}$ from Eq. (\ref{Tdot}) into (\ref{saft-1}) yields
\begin{equation}\label{saft1}
  T_{\rm A}\dot{S}_{\rm A}=\frac{\Big(f-2Tf_{\rm T}\Big)\Big(-4T^2f_{\rm TT}+Tf_{\rm T}-\frac{3}{2}f\Big)\Big(6\pi f_{\rm T}+T f_{\rm TT}(-6\pi+G\alpha)\Big)}{8\pi G T^2\Big(f_{\rm T}+2T f_{\rm TT}\Big)^2}.
\end{equation}
For $\alpha=0$, Eqs. (\ref{saft-1}) and (\ref{saft1}) reduce to Eqs. (3.7) and (3.12) of Ref. \cite{Karami:2012}, respectively. Those earlier results were obtained by neglecting the integral term when $f_{\rm TT}\ll 1$ in the entropy expression, Eq. (\ref{sat2}).

To find the matter entropy  contribution in $f(T)$ gravity, we substitute Eqs. (\ref{romT}) and (\ref{rdotA}) into Eq. (\ref{Tdsm2}) derived from the Gibbs equation (\ref{Tdsm1}). For the pressureless matter ($p_{\rm m}=0$), this gives $T_{\rm A}\dot{S}_{\rm m}$ as follows
\begin{equation}
T_{\rm
A}\dot{S}_{\rm m}=\frac{3}{4{G}T^2}\left(\frac{f-2Tf_{\rm T}}{f_{\rm T}+2Tf_{\rm TT}}\right)\Big(4T^2f_{\rm TT}-4Tf_{\rm T}+3f\Big).\label{TdsmT}
\end{equation}
By adding Eqs. (\ref{saft1}) and (\ref{TdsmT}), we obtain the expression for the GSL for any given  $f(T)$ gravity model as
\begin{eqnarray}\label{GSLt1}
&T_{\rm A}\dot{S}_{\rm
tot}=\frac{\big(f-2T f_{\rm T}\big)}{16\pi G T^2\big(f_{\rm T}+2T f_{\rm TT}\big)^2}\Bigg[3f\bigg(6\pi f_{\rm T}+T f_{\rm TT}(30\pi-G\alpha)\bigg)\nonumber\\
&+2T\bigg(-18\pi f_{\rm T}^2+4T^2 f_{\rm TT}^2(18\pi-G\alpha)+Tf_{\rm T}f_{\rm TT}(-54\pi+G\alpha)\bigg)\Bigg].
\end{eqnarray}
This expression must satisfy the condition  $T_{\rm A}\dot{S}_{\rm tot}\geq 0$.
Notably, for $\alpha=0$, the GSL equation (\ref{GSLt1})
reduces to Eq. (3.13) in Ref. \cite{Karami:2012}. This earlier result was obtained without the integral term for the apparent horizon entropy Eq. (\ref{sat2}).

When we include the integral term by setting its coefficient to $\alpha=18\pi/G$, Eqs. (\ref{saft1}) and (\ref{GSLt1}) simplify significantly
\begin{equation}\label{saft-f}
 T_{\rm A}\dot{S}_{\rm A}=\frac{3}{4{G}T^2}\left(\frac{f-2Tf_{\rm T}}{f_{\rm T}+2Tf_{\rm TT}}\right)\left(-4T^2f_{\rm TT}+Tf_{\rm T}-\frac{3}{2}f\right),
\end{equation}
\begin{equation}\label{GSLT-f}
T_{\rm A}\dot{S}_{\rm
tot}=\frac{9\left(f-2Tf_{\rm T}\right)^2}{8{G}T^2\left(f_{\rm T}+2Tf_{\rm TT}\right)}.
\end{equation}
Note that in the limit of Einstein gravity where $f(T)=T$, Eq. (\ref{GSLT-f}) reduces to
\begin{equation}
T_{\rm A}\dot{S}_{\rm tot}=\frac{9}{8{G}}>0,
\end{equation}
which shows that the GSL is always satisfied in GR.

\subsection{Results for some viable $f(T)$ models }
In this section, we analyze the revisited GSL of thermodynamics and verify the effective energy conditions for two viable $f(T)$ gravity models which justify the late time observations
\cite{Wu:2010b,linder}.
\subsubsection{Model 1}

The first $f(T)$ model is given by \cite{Wu:2010b,linder}
\begin{equation}
f(T)=T+\mu_1{(-T)}^n,~~~~~~{\rm Model~1},\label{model1}
\end{equation}
where $\mu_1$ and $n$ are model parameters. The parameter $\mu_1$, can be determined by substituting Eq. (\ref{model1}) into the first Friedmann Eq. (\ref{f1}). Solving the resulting equation for the present time gives $\mu_1=\left(\frac{1-\Omega_{\rm m_0}}{2n-1}\right)\left(6H_0^2\right)^{1-n}$, where $H_0$ and $\Omega_{\rm m_0}=(8\pi {G}\rho_{\rm m_0})/(3H_0^2)$ are the Hubble and matter density parameters at the present time.

In order to express the torsion scalar $T$ in terms of the cosmic redshift $z=a^{-1}-1$, we use $ \rho_{\rm m}$
In terms of cosmic redshift as
\begin{equation}\label{rhom0}
   \rho_{\rm m}=\rho_{\rm m_0}a^{-3}=-\frac{1}{16\pi {G}}T_0\Omega_{\rm m_0}(1+z)^{3},
\end{equation}
and substitute Eqs. (\ref{roeT}), (\ref{T}), (\ref{model1}), and (\ref{rhom0}) into the first Friedmann Eq. (\ref{f1}). This yields
\begin{equation}\label{t/t0}
 \frac{T}{T_0}=\Omega_{\rm m_0}(1+z)^3-\left(\frac{T}{T_0}\right)^{n}(\Omega_{\rm m_0}-1),~~~~~~{\rm Model~1},
\end{equation}
where $T_0=-6H_0^2$ is the current torsion scalar.

\subsubsection{Model 2}

The second  $f(T)$ model is given by \cite{Wu:2010b}
\begin{equation}
f(T)=T-\mu_2 T\Big(1-e^{\beta\frac{T_0}{T}}\Big),~~~~~~{\rm
Model~2},\label{model2}
\end{equation}
where $\mu_2$ and $\beta$ are model parameters. The parameter $\mu_2$ is obtained by replacing the $f(T)$ model (\ref{model2}) into the first Friedmann Eq. (\ref{f1}) at the present time which yields $\mu_2=\frac{1-\Omega_{\rm m_0}}{1-(1-2\beta)e^{\beta}}$.

Similar to Model 1, one can express the torsion scalar $T$ in terms of the redshift $z$ by substituting Eqs. (\ref{roeT}), (\ref{T}), (\ref{model2}) and (\ref{rhom0}) in Eq. (\ref{f1}), which gives
\begin{equation}\label{t/t02}
 \frac{T}{T_0}=\frac{\left[\left(e^{\beta\frac{T_0}{T}}-1\right)\frac{T}{T_0}-2\beta e^{\beta\frac{T_0}{T}}\right](\Omega_{\rm m_0}-1)}{1+e^{\beta}(2\beta-1)}+(1+z)^3\Omega_{\rm m_0},~~~~~~{\rm Model~2}.
\end{equation}

In the following, using Eq. (\ref{sat2}), the entropy-area relation for both $f(T)$ models (\ref{model1}) and (\ref{model2}) can be obtained as
\begin{eqnarray}
  GS_{\rm A} &\simeq& \frac{A}{4}-A^{1.96}
,~~~~~~{\rm Model~1},\\
 G S_{\rm A} &\simeq& 9.23 A+\big[7.27+A\left(45+133A\right)\big]e^{-7.41A},~~~~~~{\rm
Model~2}.
\end{eqnarray}

\begin{figure}[H]
\begin{minipage}[b]{1\textwidth}
\vspace{-.1cm}
\centering
\subfigure[\label{T1}]{\includegraphics[width=.48\textwidth]%
{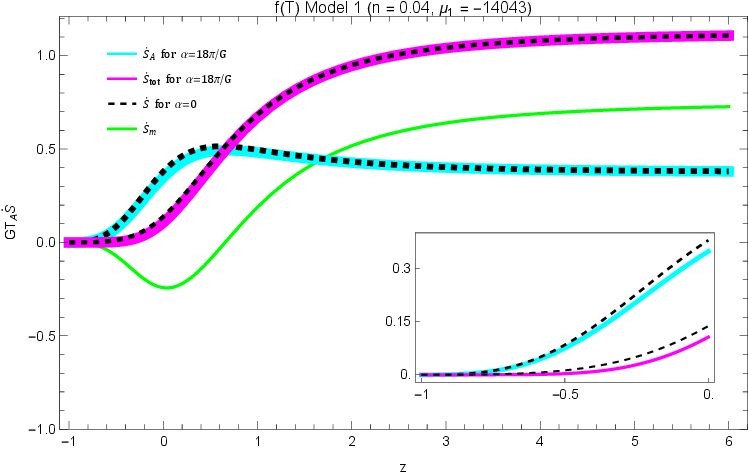}} \hspace{.1cm}
\centering
\subfigure[\label{T2}]{\includegraphics[width=.48\textwidth]%
{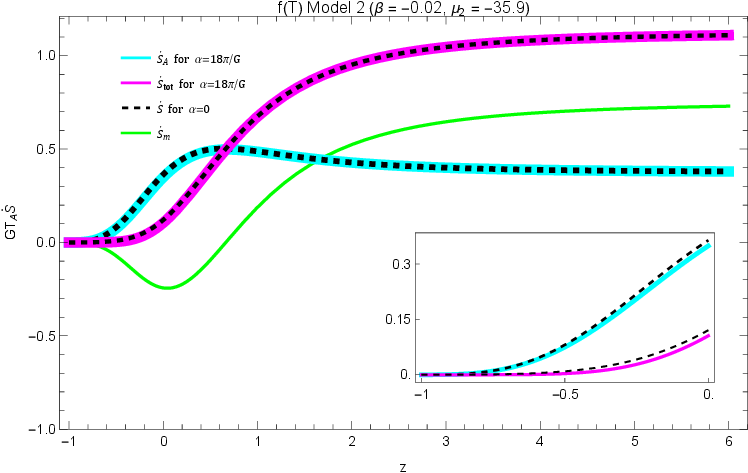}} \hspace{.1cm}
\end{minipage}
\caption{Evolutions of $T_{\rm A}\dot{S}_{\rm A}$ (\ref{saft1}), $T_{\rm A}\dot{S}_{\rm m}$ (\ref{TdsmT}), and  $T_{\rm A}\dot{S}_{\rm tot}$ (\ref{GSLt1}) as functions of the redshift $z$ for $f(T)$ gravity (a) Model 1 and (b) Model 2. Black dashed curves show the entropy rates obtained with the standard formalism ($\alpha=0$), while magenta and cyan curves denote, respectively,  the revisited total and horizon entropies including the integral term ($\alpha=18\pi/G$). The small panels highlight the late time behavior. The auxiliary parameters are $H_0=68.22~\rm km~s^{-1}~Mpc^{-1}$ and $\Omega_{\rm m_0}=0.3032$ \cite{ACT}. For Model 1 and Model 2, we set $n=0.04$ \cite{Wu:2010b,linder} and $\beta=-0.02$ \cite{Wu:2010b}, respectively. These give $\mu_1=\left(\frac{1-\Omega_{\rm m_0}}{2n-1}\right)\left(6H_0^2\right)^{1-n}=-14043$ and $\mu_2=\frac{1-\Omega_{\rm m_0}}{1-(1-2\beta)e^{\beta}}=-35.9$. }
\label{fig:fT}
\end{figure}

Next, we analyze the validity of the GSL of thermodynamics for two viable $f(T)$ gravity models (\ref{model1}) and (\ref{model2}). We compare the results obtained using the standard entropy definition ($\alpha=0$) with those from our revisited formalism, which includes the  integral term ($\alpha=18\pi/G$).
By substituting Eqs. (\ref{model1}) and  (\ref{model2}) for both $f(T)$ models into Eqs. (\ref{saft1}), (\ref{TdsmT}), and (\ref{GSLt1}) and using the redshift relations from Eqs. (\ref{t/t0}) and (\ref{t/t02}), we analyze the evolution of the  $T_{\rm A}\dot{S}_{\rm A}$ (horizon entropy), $T_{\rm A}\dot{S}_{\rm m}$  (matter entropy), and  $T_{\rm A}\dot{S}_{\rm tot}$ (total entropy or GSL) as functions of the redshift $z$, in the presence ($\alpha=18\pi/G$) and absence ($\alpha=0$) of the integral term. The numerical results are displayed in Fig. \ref{fig:fT}.

In this figure, the black dashed curves represent the entropy rates computed without the integral term ($\alpha=0$), while the magenta and cyan curves denote the revisited total and horizon entropies in the presence of the integral term ($\alpha=18\pi/G$).
It is inferable from these plots that (i)  for both models, the matter entropy violates the second law of thermodynamics, with $T_{\rm A}\dot{S}_{\rm m}<0$ in the interval $-1\leq z<0.66$, spanning  from the recent past into the future (green curves); (ii) there is only a small discrepancy between the entropy rates obtained from the standard formalism ($\alpha=0$) and revisited formalism ($\alpha=18\pi/G$) at late times, as highlighted in the zoomed panels; (iii) inclusion of the integral term in the revisited formalism ($\alpha=18\pi/G$) relatively improves the behavior of the Model 2 by removing the late time GSL violation,  while it leaves Model 1 essentially unaffected; and (iv) at late times ($z=-1$), as  $T_{\rm A}\dot{S}_{\rm tot}$ approaches zero, the universe tends asymptotically to a de Sitter phase, see Eq. (\ref{gsl}) in which $\tilde{r}_{\rm A}=H^{-1}={\rm cte.}$, $\dot{\tilde{r}}_{\rm A}=0$ and $T_{\rm A}\dot{S}_{\rm tot}=0$.
The numerical redshift intervals in which the GSL of thermodynamics is violated are summarized in Table \ref{tab:ft}.
It is deduced from this Table that including the integral term in the horizon entropy expression Eq. (\ref{sat2}), resolves the inconsistency with the GSL at the late time ($z=-1$) for Model 2, while Model 1 satisfies the GSL for both entropy prescriptions.

Furthermore, using $\rho_{\rm e}$  and $p_{\rm e}$ from Eqs. (\ref{peT})-(\ref {roeT}), along with  Eqs. (\ref{T}) and (\ref{rdotA}) for both $f(T)$ models (Eqs. (\ref{model1}) and  (\ref{model2})), and utilizing the redshift relations from Eqs. (\ref{t/t0}) and (\ref{t/t02}), we analyze the evolution of the  effective energy conditions (NEC and SEC) as functions of the redshift $z$. The numerical results are summarized in Table \ref{tab:ft}. As shown in this table, while the effective SEC is expectedly violated to allow for the accelerating expansion of the universe, the effective NEC remains robustly satisfied throughout the cosmic evolution for these models.

%
\begin{table}[h!]
\centering
	\caption{Invalid redshift interval for the GSL and the effective energy conditions in the two $f(T)$ gravity models (\ref{model1}) and (\ref{model2}). For each model, the table lists the redshift ranges where $T_{\rm A}\dot{S}_{\rm tot}<0$, Eq. (\ref{GSLt1}), for the standard entropy ($\alpha=0$) and for the revisited entropy including the integral term ($\alpha=18\pi/G$).}
	\label{tab:ft}
	{\footnotesize
		\begin{tabular}{|c  @{\hspace{8pt}}|@{\hspace{2pt}} c @{\hspace{8pt}}|@{\hspace{8pt}} c @{\hspace{8pt}}|c  @{\hspace{8pt}} |c  @{\hspace{8pt}}|}
			\thickhline
			& \multicolumn{4}{c|}{Invalid redshift interval for }  \\
			\cline{2-5}
             & GSL, Eq. (\ref{GSLt1}) &  GSL, Eq. (\ref{GSLt1})&NEC&SEC\\
			$f(T)$ Model & $\alpha=0$  & $\alpha=18\pi/G$ & $(\rho_{\rm e}+p_{\rm e})\geq 0$& $(\rho_{\rm e}+3p_{\rm e})\geq 0$\\
			\thickhline
			 Model 1 &None & None & None& All \\
\hline
			 Model 2  & \quad $z=-1$ & None  & None& All\\
			\thickhline
	\end{tabular}}
\end{table}
\section{$f(R)$ Gravity }\label{sec:fr}

Here, we are interested in examining the validity of GSL within the framework of $f(R)$ gravity with the following action
\cite{De Felice:2010,Sotiriou:2010,Nojiri:2011}
\begin{equation}\label{f action}
{\cal S} = \int {\sqrt { - g} }~{\rm d}^4
x\left[\frac{{f(R)}}{{16\pi G }} + {\cal L}_{\rm m}\right],
\end{equation}
where $g$ and ${\cal L}_{\rm m}$ denote the determinant of the metric $g_{\mu\nu}$ and the Lagrangian density of the matter, respectively. Moreover, $f(R)$ signifies an arbitrary function of the Ricci scalar $R$.

For a spatially flat FRW universe, the effective energy density and pressure owing to $f(R)$ modified gravity contributions are given by \cite{Capozziello:2003,Capozziello:2005}
\begin{eqnarray}
\rho_{\rm e} &=&\frac{1}{8\pi{G}}\left[\frac{1}{2}\big(Rf_{\rm R}-f\big)-3H\dot{f}_{\rm R}+3H^2\big(1-f_{\rm R}\big)\right],\label{ror}\\
p_{\rm e}&=&
\frac{1}{8\pi{G}}\left[\frac{-1}{2}\big(Rf_{\rm R}-f\big)+\ddot{f}_{\rm R}+2H\dot{f}_{\rm R}-(1-f_{\rm R})\big(2\dot{H}+3H^2\big)\right],\label{pr}
\end{eqnarray}
with
\begin{equation}\label{R}
R =6(\dot{H}+2H^2).
\end{equation}
Here, $f_{\rm R}\equiv{\rm d}f/{\rm d}R$.
Note that, for $f(R)=R$ in the above equations,
$\rho_{\rm e}=0=p_{\rm e}$, and the Friedmann Eqs. (\ref{f1})-(\ref{f2}) are transformed to the standard form in GR.

\subsection{Apparent horizon entropy in $f(R)$ gravity}

Adding $\rho{_e}$ and $p_{\rm e}$ from Eqs. (\ref{ror})-(\ref{pr}) results in
\begin{equation}\label{ror+pr}
\rho_{\rm e}+p_{\rm e}=\frac{1}{8\pi{G}}\Big[\ddot{f}_{\rm R}-H\dot{f}_{\rm R}-2\dot{H}(1-f_{\rm R})\Big].
\end{equation}

Now, substituting $\rho_{\rm e}+p_{\rm e}$ from the above equation into the universal relation for the apparent horizon entropy (\ref{sa}) gives rise to
\begin{equation}\label{sar1}
  S_{\rm A}=\frac{A}{4{G}}-\frac{\pi}{G}\int H\tilde{r}^4_{\rm A}\left[\ddot{f}_{\rm R}-H\dot{f}_{\rm R}-2\dot{H}\left(1-f_{\rm R}\right)\right]{\rm d}t.
\end{equation}
Here, using $H\tilde{r}_{\rm A}=1$ and $\dot {H}=-\dot{\tilde{r}}_{\rm A}/\tilde{r}^2_{\rm A}$ we have
\begin{equation}\label{sar2}
  S_{\rm A}=\frac{A}{4{G}}-\frac{\pi}{G}\Bigg(\int \ddot{f}_{\rm R}\tilde{r}^3_{\rm A}{\rm d}t-\int \tilde{r}^2_{\rm A}{\rm d}f_{\rm R}+\int 2\tilde{r}_{\rm A}{\rm d}\tilde{r}_{\rm A}-\int 2f_{\rm R}\tilde{r}_{\rm A}{\rm d}\tilde{r}_{\rm A}\Bigg).
\end{equation}
Now, using the integration by part for the second integration term, $\int \tilde{r}^2_{\rm A}{\rm d}f_{\rm R}=\tilde{r}^2_{\rm A}f_{\rm R}-\int f_{\rm R}{\rm d}\left(\tilde{r}^2_{\rm A}\right)$, results in
\begin{equation}\label{sar3}
  S_{\rm A}=\frac{A f_{\rm R}}{4{G}}-\beta\int \ddot{f}_{\rm R}\tilde{r}^3_{\rm A}{\rm d}t,
\end{equation}
where $\beta\equiv\frac{\pi}{G}$ and we have used $-\frac{\pi}{G}\int 2\tilde{r}_{\rm A}{\rm d}\tilde{r}_{\rm A}=-\frac{\pi}{G}\tilde{r}_{\rm A}^2=-\frac{A}{4{G}}$. It is notable that the horizon entropy in $f(R)$ gravity obtained from Eq. (\ref{sar3}) has a modified integral term compared to the common version $S_{\rm A}=Af_{\rm R}/(4{G})$
used in literature \cite{Wald}.
\subsection{GSL in $f(R)$ gravity}

In what follows, the impact of the integral term on the horizon entropy (\ref{sar3}) and thereby the GSL of thermodynamics in $f(R)$ gravity is recomputed. First, using Eq. (\ref{TA1}) and time-variation of Eq. (\ref{sar3}), $T_{\rm A}\dot{S}_{\rm A}$ is calculated as follows
\begin{equation}\label{tsra-1}
T_{\rm A}\dot{S}_{\rm A} =\frac{1}{4GH} \left( 2-\dot{\tilde{r}}_{\rm A} \right) \left(
\frac{2\dot{\tilde{r}}_{\rm A}}{\tilde{r}_{\rm A}}f_{\rm R}+\dot{f}_{\rm R}-\frac{\beta G}{\pi}\tilde{r}_{\rm A}\ddot{f}_{\rm R}\right),
\end{equation}
which, using $\tilde{r}_{\rm A}=H^{-1}$ and $\dot{\tilde{r}}_{\rm A}=-\dot{H}H^{-2}$, turns to
\begin{equation}\label{tsra-2}
T_{\rm A}\dot{S}_{\rm A} =\frac{1}{4GH^4}\left(\dot{H}+2H^2\right)\left(-2\dot{H}f_{\rm R}+H\dot{f}_{
\rm R}-\frac{\beta G}{\pi}\ddot{f}_{\rm R}\right).
\end{equation}
At this stage, we want to calculate the matter contribution to the total entropy. To do so, by substituting Eqs. (\ref{ror})-(\ref{R}) into modified Friedmann equations (\ref{f1})-(\ref{f2}), we can compute $\rho_{\rm m}+p_{\rm m}$ in $f(R)$ gravity as follows
\begin{equation}\label{romr}
  \rho_{\rm m}+p_{\rm m}=\frac{1}{8\pi {G}}\left[-2\dot{H}f_{\rm R}+H\dot{f}_{\rm R}-\ddot{f}_{\rm R}\right].
\end{equation}
Here, $\rho_{\rm m}$ and $p_{\rm m}$ denote the physical (minimally coupled) matter components, which obey the conservation law Eq. (\ref{clm}).  Substituting the combination $\rho_{\rm m}+p_{\rm m}$ from Eq. (\ref{romr}) into the Gibbs-based relation Eq. (\ref{Tdsm2}), and using $\dot{\tilde{r}}_{\rm A}=-\dot{H}H^{-2}$ and $\tilde{r}_{\rm A}=H^{-1}$, we obtain  the matter contribution to the entropy rate as follows
\begin{equation}\label{smr}
  T_{\rm A}\dot{S}_{\rm m}=\frac{1}{4{G}H^4}\Big(-2\dot{H} -2H^2\Big)\left(-2\dot{H}f_{\rm R}+H\dot{f}_{\rm R}-\ddot{f}_{\rm R}\right).
\end{equation}
Finally, adding Eqs. (\ref{tsra-2}) and (\ref{smr}) gives rise to the revisited GSL of thermodynamics in $f(R)$ gravity as follows
\begin{equation}\label{gslr1}
 T_{\rm A}\dot{S}_{\rm tot}=\frac{1}{4GH^{4}}\left[2\dot{H}^2f_{\rm R}-H\dot{H}\dot{f}_{\rm R}+\ddot{f}_{\rm R}\left(-\frac{\beta G}{\pi}\big(\dot{H}+2H^2\big)+2\big(\dot{H}+H^2\big)\right)\right].
\end{equation}
For $\beta=0$, the equation (\ref{tsra-2}) and GSL equation (\ref{gslr1}), respectively
reduce to Eqs. (29) and (30) of Ref. \cite{Asadzadeh:2016}, which were obtained in the absence of the integral term for the apparent horizon entropy Eq. (\ref{sar3}). Considering the integral term by setting  $\beta=\pi/G$, Eqs. (\ref{tsra-2}) and (\ref{gslr1}) simplify to
\begin{equation}\label{tsra-3}
T_{\rm A}\dot{S}_{\rm A} =\frac{1}{4GH^4}\left(\dot{H}+2H^2\right)\left(-2\dot{H}f_{\rm R}+H\dot{f}_{
\rm R}-\ddot{f}_{\rm R}\right),
\end{equation}
\begin{equation}\label{gslr}
 T_{\rm A}\dot{S}_{\rm tot}=\frac{\dot{H}}{4GH^{4}}\left[2\dot{H}f_{\rm R}-H\dot{f}_{\rm R}+\ddot{f}_{\rm R}\right].
\end{equation}
To satisfy the GSL of thermodynamics in $f(R)$ gravity, the above equation should be amenable to $T_{\rm A}\dot{S}_{\rm tot}\geq 0$. Note that, for a de Sitter universe ($\dot{H}=0$), the GSL equation (\ref{gslr}) reduces to $T_{\rm A}\dot{S}_{\rm tot}=0$ which corresponds to an adiabatic universe. In the case of Einstein gravity $f(R)=R$, the GSL equation (\ref{gslr}) simplifies to
\begin{equation}\label{gslr-gr}
 T_{\rm A}\dot{S}_{\rm tot}=\frac{\dot{H}^2}{2{G}H^4}=\frac{9}{8G}>0,
\end{equation}
where we have used the Friedmann Eqs. (\ref{f1})-(\ref{f2}) with $K=0$, $\rho_{\rm e}=0=p_{\rm e}$ and $p_{\rm m}=0$. This indicates that the GSL is always satisfied in GR.

\subsection{Results for $f(R)$ Models }
In the following, we are interested in re-examining the validity of GSL of thermodynamics, using our general formalism, in five viable $f(R)$ gravity models, consisting of the AB, Starobinsky, Exponential, Hu-Sawicki, and  Tsujikawa models. It should be noted that all of these models were previously evaluated using the conventional formulation for the horizon entropy, $S_{\rm A}=Af_{\rm R}/(4{G})$, without the integral term in (\ref{sar3}), as presented in \cite{Asadzadeh:2016}. Moreover we evaluate the effective energy conditions for all of the $f(R)$models.

\subsubsection{AB Model}
The AB $f(R)$ model is defined by \cite{AB,AB1,AB2}
\begin{equation}\label{AB1}
    f(R)=\frac{R}{2}+\frac{\epsilon}{2}\log\left[\frac{\cosh\big(\frac{R}{\epsilon}-b\big)}{\cosh(b)}\right],
\end{equation}
wherein $b=1.4$ is a dimensionless parameter, and $\epsilon=R_{\rm
s}/\big[b+\log(2\cosh b)\big]$. Following \cite{Asadzadeh:2016},   $ R_{\rm s}= -36\;\Omega_{\rm
m_0}H^2_0\big[b+\log(2\cosh b)\big]/{\log\big(\frac{1-\tanh
b}{2}\big)}$ is obtained in the high curvature domain when the behavior of the AB $f(R)$ model
(\ref{AB1}) becomes similar to the $\Lambda$CDM model with
$f(R)=R-2\Lambda$.
\subsubsection{Starobinsky Model}
The Starobinsky $f(R)$ model is given by \cite{Starobinsky}
\begin{equation}\label{Staro}
    f(R)=R+\lambda R_{\rm
s}\left[\left(1+\frac{R^2}{R^2_{\rm s}}\right)^{-n}-1\right],
\end{equation}
where $n=2$ and $\lambda=1$ are dimensionless parameters  \cite{Motohashi}. In addition $R_{\rm s}=18\Omega_{\rm m_0}H^2_0/\lambda$ is obtained at early times (large redshift regime), when the Starobinsky $f(R)$ model behaves like the $\Lambda$CDM model with $f(R)=R-2\Lambda$.
\subsubsection{Exponential Model}
The Exponential $f(R)$ model is introduced by \cite{Bamba:2010,Elizalde:2011,Cognola:2008},
\begin{equation}\label{EXP}
    f(R)=R-\gamma R_{\rm s}\left(1-e^{-\frac{R}{R_{\rm s}}}\right),
\end{equation}
where  $\gamma=1.8$ \cite{Bamba:2010} is a dimensionless parameter, and $R_{\rm
s}=18\Omega_{\rm m_0}H^2_0/\beta$ denotes the curvature modification
scale.
\subsubsection{Hu-Sawicki Model}
The Hu-Sawicki $f(R)$ model is  defined by \cite{HS}
\begin{equation}\label{HS}
    f(R)=R-\frac{c_1 R_{\rm s}\big(\frac{R}{R_{\rm s}}\big)^n}{c_2\big(\frac{R}{R_{\rm s}}\big)^n+1},
\end{equation}
where $n=4$, $c_1=1.25\times10^{-3},
c_2=6.56\times10^{-5}$ \cite{Luisa}, and  $R_{\rm
s}=18c_2\Omega_{\rm m_0 }H^2_0/c_1$ are constant parameters of the model.
\subsubsection{Tsujikawa Model}
The Tsujikawa $f(R)$ model is identified as \cite{Tsujik}
\begin{equation}\label{Tsu}
    f(R)=R-\lambda R_{\rm s} \tanh{\left(\frac{R}{R_{\rm s}}\right)},
\end{equation}
where $\lambda=1$ \cite{Bamba2,Bamba3}, and $R_{\rm s}$ is obtained as $R_{\rm s}=18\Omega_{\rm m_0}H^2_0/\lambda$.

\begin{figure}[H]
\begin{minipage}[b]{1\textwidth}
\vspace{-1.cm}
\centering
\subfigure[\label{AB}]{\includegraphics[width=.47\textwidth]%
{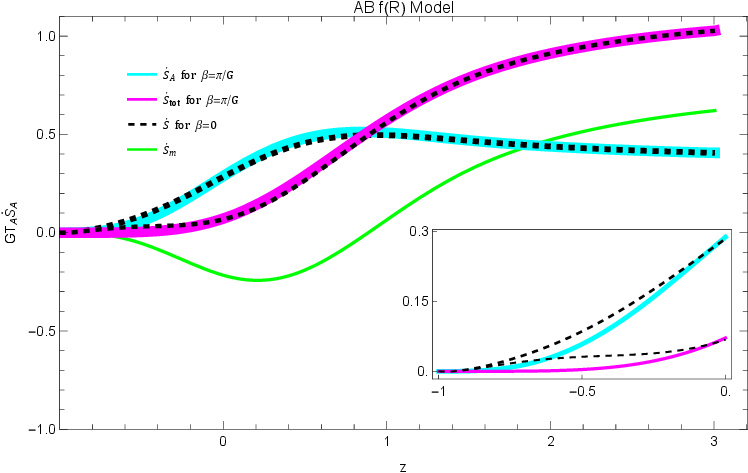}} \hspace{.1cm}
\centering
\subfigure[\label{St}]{\includegraphics[width=.47\textwidth]%
{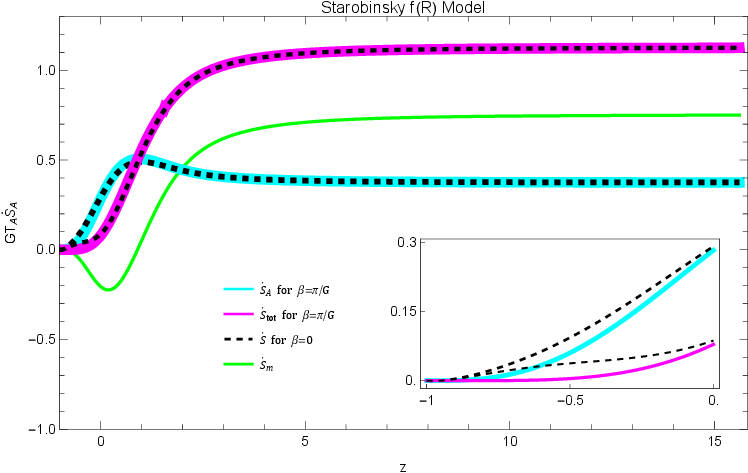}} \hspace{.1cm}
\centering
\subfigure[\label{Ex}]{\includegraphics[width=.47\textwidth]%
{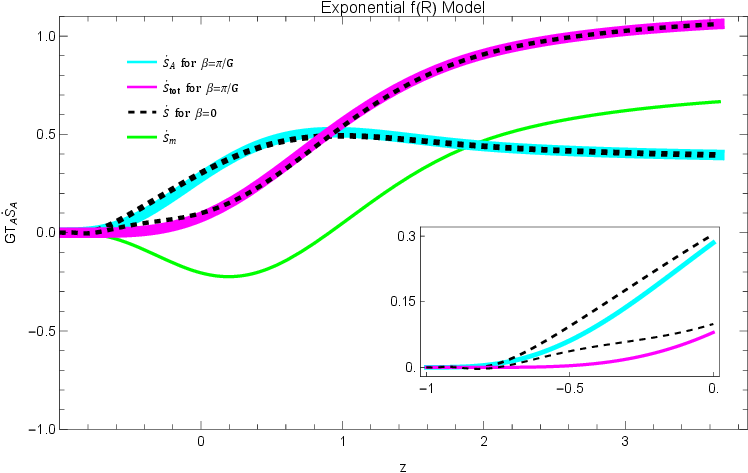}} \hspace{.1cm}
\centering
\subfigure[\label{Hu}]{\includegraphics[width=.47\textwidth]%
{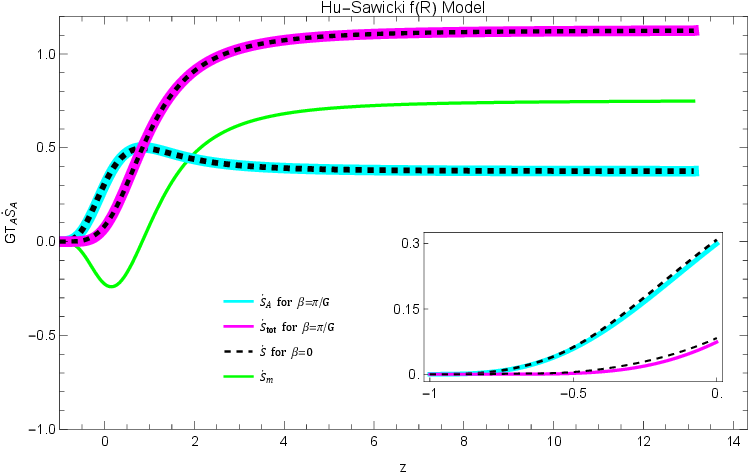}} \hspace{.1cm}
\centering
\subfigure[\label{Ts}]{\includegraphics[width=.47\textwidth]%
{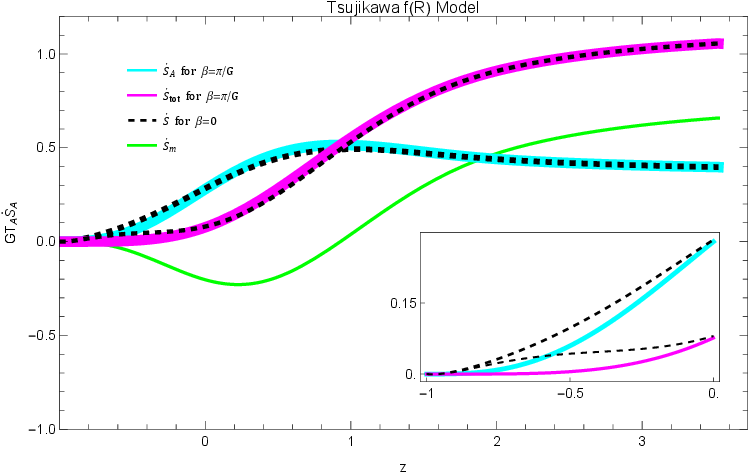}} \hspace{.1cm}
\end{minipage}
\vspace{-0.5cm}
\caption{Evolutions of $T_{\rm A}\dot{S}_{\rm A}$ (\ref{tsra-2}),  $T_{\rm A}\dot{S}_{\rm m}$ (\ref{smr}), and  $T_{\rm A}\dot{S}_{\rm tot}$ (\ref{gslr1}) versus the redshift $z$ for (a) AB, (b) Starobinsky, (c) Exponential, (d) Hu--Sawicki, and (e) Tsujikawa \(f(R)\) models. Black dashed curves show the entropy rates in the standard formalism without the integral term ($\beta=0$), whereas magenta and cyan curves denote, respectively, the revisited total and horizon entropies including the integral term ($\beta=\pi/G$). The insets highlight the late time domain. The auxiliary parameters are $H_0=68.22~\rm km~s^{-1}~Mpc^{-1}$ and $\Omega_{\rm m_0}=0.3032$ \cite{ACT}.}
\label{fig:fR}
\end{figure}

At this stage, we examine the apparent horizon entropy, matter entropy, and total entropy in their revisited forms, given by Eqs. (\ref{tsra-2}), (\ref{smr}), and (\ref{gslr1}) for the $f(R)$ gravity models introduced above. We study the effect of including the integral term in the horizon and total entropies ($\beta=\pi/G$) in comparison with the standard expressions without this term ($\beta=0$). To this end, following the numerical procedure developed in Ref. \cite{Asadzadeh:2016}, we substitute the $f(R)$ functions of Eqs.  (\ref{AB1})-(\ref{Tsu}) into Eqs. (\ref{tsra-2}), (\ref{smr}), and (\ref{gslr1}) to obtain the evolution of the $T_{\rm A}\dot{S}_{\rm A}$ (horizon entropy), $T_{\rm A}\dot{S}_{\rm m}$  (matter entropy), and  $T_{\rm A}\dot{S}_{\rm tot}$ (total entropy or GSL). The numerical results are displayed in Fig. \ref{fig:fR}. In this figure, similarly to Fig. \ref{fig:fT}, the black dashed curves illustrate the entropy rates from the standard approach without the integral term ($\beta=0$), whereas the magenta and cyan curves represent, respectively, the revisited total and horizon entropies including the integral term ($\beta=\pi/G$).
From this figure one can find that (i) for all $f(R)$ models, the matter entropy violates the second law of thermodynamics, with $T_{\rm A}\dot{S}_{\rm m}<0$ from the near past ($z\simeq 0.9$) to the future ($z=-1$) (green curves); (ii) in the redshift interval $z\in[-1,0]$, the entropy rates obtained in the revisited formalism with $\beta=\pi/G$ (magenta and cyan curves)  deviate from those of the standard formalism with $\beta=0$  (black dashed curves), while at earlier times the curves almost coincide (see the zoomed panels).
The redshift intervals in which the GSL and the effective energy conditions are violated for these $f(R)$ models are summarized in Table \ref{tab:fr}. The Table shows that including the integral term in revisited formalism improves relatively the late time consistency of the Exponential and Tsujikawa $f(R)$ models with the GSL, while it has no effect on the remaining models.

Finally, using $\rho_{\rm e}$ and $p_{\rm e}$ from Eqs (\ref{ror}) and (\ref{pr}), we evaluate the effective energy conditions (NEC and SEC) for all $f(R)$ models. As summarized in Table \ref{tab:fr}, these models exhibit a more complex thermodynamic behavior: while the effective SEC is violated across the entire cosmic evolution to drive late-time acceleration, the effective NEC is violated within specific redshift intervals. This broken NEC provides a critical opportunity to test the robustness of the revisited GSL formula. Remarkably, as observed in our analysis, the GSL ($T_{\rm A}\dot{S}_{\rm tot}\ge0$) remains satisfied even in regimes where standard energy conditions fail, thereby reinforcing the thermodynamic consistency of the revisited entropy formalism in modified gravity scenarios.

\begin{table}[h!]
\centering
	\caption{Invalid redshift intervals for the GSL and the effective energy conditions in the $f(R)$ gravity models, Eqs.  (\ref{AB1})-(\ref{Tsu}). For each model, the table lists the redshift ranges where $T_{\rm A}\dot{S}_{\rm tot}<0$, Eq. (\ref{gslr1}), for the standard entropy without the integral term ($\beta=0$) and for the revisited entropy including the integral term ($\beta=\pi/G$).}
	\label{tab:fr}
	{\footnotesize
		\begin{tabular}{|c  @{\hspace{1pt}}|@{\hspace{1pt}} c @{\hspace{1pt}}|@{\hspace{1pt}} c @{\hspace{1pt}}|c  @{\hspace{1pt}}|c  @{\hspace{1pt}}|}
			\thickhline
			& \multicolumn{4}{c|}{Invalid redshift intervals for }  \\
			\cline{2-5}
             & GSL, Eq. (\ref{gslr1}) &  GSL, Eq. (\ref{gslr1})&NEC&SEC\\
			$f(R)$ Model & $\beta=0$  & $\beta=\pi/G$ & $(\rho_{\rm e}+p_{\rm e})\geq 0$&$(\rho_{\rm e}+3p_{\rm e})\geq 0$\\
			\thickhline
			AB  &$[-1,-0.94]$ & $[-1,-0.94]$& $[-1,-0.94]$,$[0.31,2.61]$& All \\
            \hline
             Starobinsky&$[-1,-0.96]$ & $[-1,-0.96]$& $[-1,-0.95]$,$[0.67,8.14]$ & All\\
            \hline
			Exponential&$[-1,-0.96]$,$[-0.89,-0.75]$ & $[-1,-0.96]$,$[-0.87,-0.80]$& $[-0.89,-0.74]$,$[0.58,3.38]$ & All \\
            \hline
            Hu-Sawicki& None & None& $[0.64,6.51]$,$[7.07,8.25]$,& \\
            & & &[10.81,12.25] & All\\
            \hline
             Tsujikawa&$[-1,-0.96]$ & None& $[-1,-0.95]$,$[0.48,3.25]$ & All\\
			\thickhline
	\end{tabular}}
\end{table}

%
%
%
%
\section{Conclusions}{\label{sec:con}}
Here, a universal and revisited formalism for the apparent horizon entropy in modified gravity was constructed by combining the Clausius relation with the modified Friedmann equations in a FRW universe.
This establishes the effective connection between gravity and thermodynamics in the context of modified gravity theories. The revisited and universal expression for the entropy of the apparent horizon was derived from the reformulated Friedmann equations, which includes an integral term dependent on the effective energy density and pressure arising from the gravitational modifications, Eq. (\ref{sa}). This approach ensures the equivalence between the Clausius relation and the gravitational field equations for any modified gravity theory. On this basis, a universal expression for the generalized second law (GSL) was obtained, which is applicable to any modified gravity model.

Thereafter, the revisited methodology was applied to re-examine the GSL for some viable $f(T)$ and $f(R)$ gravity models that have been previously studied, which give rises to a direct comparison between the standard entropy prescriptions and the revisited ones including the integral term appeared in the apparent horizon entropy. For the $f(T)$ case, the Model 1 was found to satisfy the GSL in both prescriptions, while the Model 2 exhibits a late time GSL violation when the integral term is neglected but becomes thermodynamically consistent once the correction is included. For the $f(R)$ sector, the integral term leaves the GSL behavior of the AB, Starobinsky, and Hu–Sawicki models essentially unchanged, but improves relatively the late-time GSL satisfaction in the Exponential and Tsujikawa models. Analysis illustrates that, through our revisited apparent horizon entropy, the GSL is robustly satisfied throughout the cosmic evolution for all $f(T)$ models (Table \ref{tab:ft}) and two $f(R)$ models such as Tsujikawa and Hu-Sawicky (Table \ref{tab:fr}).

These results indicate that the integral contribution to the apparent horizon entropy is not merely a technical refinement but can play an amending role in the thermodynamic viability of specific modified gravity scenarios, particularly near the asymptotic de Sitter regime. The universal entropy and GSL relations developed here thus provide a unified and robust framework for testing the consistency of cosmological models with horizon thermodynamics and can be straightforwardly extended to other classes of modified gravity or to scenarios including additional cosmic components and interactions.

\subsection*{Acknowledgements}
The authors thank the referee for his/her valuable comments.

\end{document}